\documentstyle[prb,aps,epsf,floats]{revtex}
\begin{document}
\renewcommand{\textfraction}{0.0}
\renewcommand{\floatpagefraction}{.7}
\setcounter{topnumber}{5}
\renewcommand{\topfraction}{1.0}
\setcounter{bottomnumber}{5}
\renewcommand{\bottomfraction}{1.0}
\setcounter{totalnumber}{5}
\setcounter{dbltopnumber}{2}
\renewcommand{\dbltopfraction}{0.9}
\renewcommand{\dblfloatpagefraction}{.7}
\draft

\twocolumn[\hsize\textwidth\columnwidth\hsize\csname@twocolumnfalse%
\endcsname

\title{Study of Chirality in the Two-Dimensional XY Spin Glass}

\author{H. S. Bokil and A. P. Young}
\address{Department of Physics, University of California, Santa Cruz, CA 95064}


\maketitle

\begin{abstract} We study the chirality in the Villain form of the XY
spin glass in two--dimensions by Monte Carlo simulations. We calculate
the chiral-glass correlation length exponent $\nu_{\scriptscriptstyle
CG}$ and find that $\nu_{\scriptscriptstyle CG} = 1.8 \pm 0.3$ in
reasonable agreement with earlier studies. This indicates that the
chiral and phase variables are decoupled on long length scales and
diverge as $T \to 0$ with {\em different} exponents, since the
spin-glass correlation length exponent was found, in earlier studies,
to be about 1.0.
\end{abstract}
\vskip 1cm ]

Ever since Villain\cite{villain} pointed out the existence of a discrete
reflection symmetry (in
addition to the ordinary rotation)
in frustrated vector spin systems, there has been considerable interest in
the similarity or difference between the behavior the
variables corresponding to these two symmetries -
chiralities (which are quenched in vortices), corresponding to reflection,
and spins,
corresponding to rotation.
In part this interest comes from the observation that spin glasses seem
to be in the Ising universality class, though many of them should be
described quite well by a Heisenberg model.
This has led to speculations
that chiralities and spins order differently and that the Ising
behavior seen in experiments might indicate the existence of
a chiral glass phase in the absence of spin glass
ordering\cite{kawamura}. This view is supported by some numerical
simulations\cite{kt,rm}. 
However, in spite of many studies, the
problem still remains controversial.

Kawamura and Tanemura\cite{kt} studied the two dimensional XY spin 
glass by a
domain wall renormalization group technique and were the first to
present evidence that the chiral and spin glass correlation length
exponents are different in two-dimensions. They also reported Monte
Carlo simulations which supported their claim\cite{kt}. A little
later, Ray and Moore\cite{rm} also reported Monte Carlo results
which indicated that in two-dimensions the chiral and spin glass
correlation length exponents were indeed different. They estimated
$\nu_{\scriptscriptstyle SG} \simeq 1.0$ and
$\nu_{\scriptscriptstyle CG} \simeq 2.0$, the former result being
in good agreement with earlier work of Jain and Young~\cite{jy}.
More recently, Kawamura\cite{k} reported Monte Carlo simulations 
of the three
dimensional XY spin glass and claimed that there is indeed a stable
chiral glass phase. This would appear to have settled the issue.
However, recent analytic work on the
one-dimensional ladder lattice\cite{nhm} 
and on the two-dimensional system
with a special choice of disorder\cite{nh}
points in the opposite direction.
Given this controversy it seems a reasonable time to study the
two-dimensional system once again numerically.

The earlier work used a representation of the model
in terms of the phases of the
XY model, see Eq.~(\ref{ham}) below,
and vortices are expressed in terms of correlations of the
phases around an elementary square. However, vortices are only well
defined when the nearest-neighbor spin-spin correlation function is
large, which means that the temperature is already quite low. It is
therefore difficult to study vortex correlations over a
large range of temperature. In our work, we use a different
representation of the model, expressed directly in terms of the vortices
themselves. As a result it is possible to study the vortex correlations
over a larger range of temperature than before. 

The model used in the simulations is
an XY spin glass in which the interactions, $J_{ll^\prime}$,
have values $\pm J$. The Hamiltonian is
\begin{equation}
{\cal H} = - \sum_{\langle l,l^\prime\rangle}
J_{ll^\prime} \vec{S}_l \cdot \vec{S}_{l^\prime} \ ,
\label{xyham}
\end{equation}
where the $\vec{S}_i$ are two component vectors of unit length.
This can also be written as
\begin{equation}
{\cal H} = - J \sum_{\langle l,l^\prime\rangle}
\cos ( \phi_l - \phi_{l^\prime} - A_{ll^\prime}), 
\label{ham}
\end{equation}
where
$\phi_l$ is the angle (phase) the XY spin makes with a fixed direction,
and the $A_{ll^\prime}$ take
values 0 and $\pi$ with equal probability, corresponding to
$J_{ll^\prime}=1$ and $-1$ respectively.
We take the sites, $l$, to
lie on a square lattice
of size $L\times L$,
and the interaction is between all nearest neighbor pairs,
counted once.

As discussed above, it is easier to study the model in terms of vortices.
To do so we first replace the cosine in Eq.~(\ref{ham}) by the Villain
periodic Gaussian function, i.e.
\begin{equation}
{\cal H} = \sum_{\langle l,l^\prime\rangle}
V( \phi_l - \phi_{l^\prime} - A_{ll^\prime}), 
\label{ham_vill}
\end{equation}
where
\begin{equation}
\exp\left({-V(x) \over T} \right) =
\sum_{m=-\infty}^\infty
\exp\left({-J(x - 2 \pi m)^2 \over 2 T}\right) \ ,
\end{equation}
in units where Boltzmann's constant is unity. 
Performing
standard manipulations\cite{vill,jkkn}
one finds that the partition function of the Hamiltonian in
Eq.~(\ref{ham_vill}) is the same (apart from an unimportant smoothly
varying prefactor) as that of the following Hamiltonian,
\begin{equation}
\label{vortex}
{\cal H}_V = -{1\over 2} \sum_{i,j}
(n_i - b_i)G(i - j) (n_j - b_j)  ,
\label{ham_vortex}
\end{equation}
where the vortices $\{n_i\}$ run over all integer values, subject to the
``charge neutrality'' constraint
\begin{equation}
\sum_i n_i = 0 \ ,
\label{ch-neut}
\end{equation}
and $G(i - j)$ is the vortex interaction, 
\begin{equation}
\label{vortex_int}
{G(i - j) \over (2 \pi)^2} = 
{J \over N } \sum_{{\bf k} \ne 0}
{1 - \exp[i {\bf k} \cdot ({\bf r_i} - {\bf r_j})] \over
4 - 2 \cos k_x - 2 \cos k_y } \quad .
\end{equation}
Note that the Fourier transform of the
vortex interaction is $\sim k^{-2} $ for small $k$ which corresponds to
a long range logarithmic interaction in real space. 
The vortices sit on the sites $i$ of the dual lattice
which are in the centers of the squares of the original lattice.
The $b_i$ are given by (1/$2 \pi$) times
the directed sum of the quenched random variables, $A_{ll^\prime}$, on
the links of the original lattice which surround the site $i$ of the
dual lattice. They satisfy a constraint similar to
Eq.~(\ref{ch-neut}),
\begin{equation}
\sum_i b_i = 0 \ .
\end{equation}
From now on we set the interaction strength, $J$, to be unity. 

For $A_{ll^\prime} = 0$ or $\pi$ there are two kinds of
plaquettes on the original lattice,
or equivalently two kinds of sites on the
dual lattice:
there are unfrustrated sites, on which $b_i$ is integer, and 
frustrated sites, on which $b_i$ is half-integer. If $A_{ll^\prime}$ has
equal probability to takes values $0$ and $\pi$ then half the sites on
the dual lattice will be frustrated and half unfrustrated on average. 
In the ground state, most unfrustrated sites will have $n_i=0$, while
most frustrated sites will have $n_i = \pm 1/2$. 
These Ising-like variables 
are precisely the chirality variables of Villain.
It seems reasonable that the essential physics
will be preserved if one just keeps these chirality variables, i.e.
one fixes $n_i$ to be zero
on the unfrustrated sites and only allows $n_i$
to be $\pm 1/2$ on the frustrated sites. Thus one has an Ising model
with a long range antiferromagnetic coupling,
\begin{equation}
\label{chiral}
{\cal H}_{\scriptscriptstyle CG} = -{1\over 2} \sum_{i,j}
\tilde{G}(i - j) \sigma_i \sigma_j \epsilon_i \epsilon_j  \, 
\end{equation}
where, for convenience, we represent
the chiralities by Ising spins,
$\sigma_i$, of {\em unit} length so
$\sigma_i = \pm 1$, and the interaction, $\tilde{G}$, is
then equal to $G/4$. The quenched variable
$\epsilon_i$ is equal to 1 if there is a chirality on site $i$, and
is equal to 0 if there is no chirality. The lattice sites with
chirality are to be chosen at random with 50\% probability.
We impose the additional
contraint that the total number of chiralities
is {\em exactly} half the number
of lattice sites. Defining $N$ to be the number of chiralities
we have, for {\em every} sample,
\begin{equation}
N = {L^2 \over 2} .
\end{equation}

We now discuss the scaling theory and the details of our Monte Carlo 
simulations. We run two independent replicas of the
system in parallel with the same realization of the disorder and compute
the overlap between the states in the two replicas,
\begin{equation}
\label{q_replica}
q = {1 \over N} \sum_{i=1}^N \sigma_i^{(1)} \sigma_i^{(2)} \ .
\end{equation}
Here, and in the rest of the paper, the sum over sites $i$ and $j$
on the dual lattice
is only over those sites occupied by a chirality. 
The standard spin glass order parameter is just $[\langle q\rangle]$,
where the angular brackets denote the thermal averages and the square
brackets the average over disorder. 
Two useful quantities are the Binder moment ratio
$g_{\scriptscriptstyle CG}$ and the chiral glass
susceptibility $\chi_{\scriptscriptstyle CG}$ defined by
\begin{equation}
\label{moment}
g_{\scriptscriptstyle CG} = {1 \over 2}
\left\{3 - {[\langle q^4 \rangle] \over [\langle q^2 \rangle]^2}\right\}
\label{g-def}
\end{equation}
and
\begin{eqnarray}
\chi_{\scriptscriptstyle CG} & = & {1 \over N} \sum_{i,j}
[\langle \sigma_i \sigma_j \rangle^2 ] \nonumber \\
 & = & N [\langle q^2 \rangle]  \ .
\label{chi-def}
\end{eqnarray} 
$g_{\scriptscriptstyle CG}$ is defined so that it tends to 0 at high
temperatures in the thermodynamic limit,
and tends to unity as $T\to 0$ if the ground state is
non-degenerate. 
The chiral glass susceptibility should be contrasted with the spin glass
susceptibility, $\chi_{\scriptscriptstyle SG}$, defined by
\begin{eqnarray}
\chi_{\scriptscriptstyle SG} & = & {1 \over L^2} \sum_{l,l^\prime}
[\langle \cos(\theta_l - \theta_{l^\prime}) \rangle^2] \nonumber \\
 & = & L^2 \sum_{\alpha,\beta}
[\langle q_{\alpha\beta}^2 \rangle] \ ,
\label{chi-sg}
\end{eqnarray}
where
\begin{equation}
q_{\alpha\beta} = {1 \over L^2} \sum_{l=1}^{L^2}
S_{l,\alpha}^{(1)} S_{l,\beta}^{(2)} \ .
\end{equation}
Here $\alpha$ and $\beta$ denote the
components of the XY spins in Eq.~(\ref{xyham}) and
take values $x$ and $y$.

Since the transition in this system is expected to be at $T = 0$ the
following scaling forms are expected for
$\chi_{\scriptscriptstyle CG}$
and for $g_{\scriptscriptstyle CG}$\cite{by}:
\begin{equation}
\label{g_scale}
g_{\scriptscriptstyle CG} = \tilde{g}_{\scriptscriptstyle CG}
(L^{1/\nu_{_{\!CG}}} T)
\end{equation}
and
\begin{equation}
\label{chi_scale}
\chi_{\scriptscriptstyle CG} = L ^ {2 - \eta_{_{\!CG}}}
\tilde{\chi}_{\scriptscriptstyle CG} (L^{1/\nu_{_{\!CG}}} T) \ .
\end{equation}
Here the exponent $\nu_{\scriptscriptstyle CG}$ is the
correlation length exponent and $\eta_{\scriptscriptstyle CG}$ is
related to the ground state degeneracy---if the ground state is unique,
one expects $\eta_{\scriptscriptstyle CG} = 0$.
At a critical point, the data for $g_{\scriptscriptstyle CG}$ 
should be independent of size. This is a particularly
convenient way of locating the transition. The spin glass
susceptibility in Eq.~(\ref{chi-sg}) has the finite size scaling
form
\begin{equation}
\chi_{\scriptscriptstyle SG} = L ^ {2 - \eta_{_{\!SG}}}
\tilde{\chi}_{\scriptscriptstyle SG} (L^{1/\nu_{_{\!SG}}} T) \ ,
\end{equation}
with exponents $\nu_{\scriptscriptstyle SG}$ and
$\eta_{\scriptscriptstyle SG}$ which are expected to be
{\em different} from the corresponding chiral glass exponents. In
this paper, we just focus on the chiral glass critical behavior.

We use standard methods\cite{by} to ensure equilibration 
of the Monte Carlo simulation. Various quantities are computed
both from
overlaps between the two replicas and from a single replica at different
times, see Bhatt and Young\cite{by} for details. 
Typically, for the largest lattice sizes we needed
about 100000 Monte Carlo sweeps for equilibration at the lowest
temperature. For the averaging
over disorder we took between 1000 and 10000 samples. For most
of the data points the
statistical errors were estimated by averaging over all the
samples. However, for the largest lattice sizes and lowest temperatures
we divided up the samples into blocks of a few hundred each and
calculated the errors from the standard deviations of the
quantities between different blocks.
\begin{figure}
\epsfxsize=\columnwidth\epsfbox{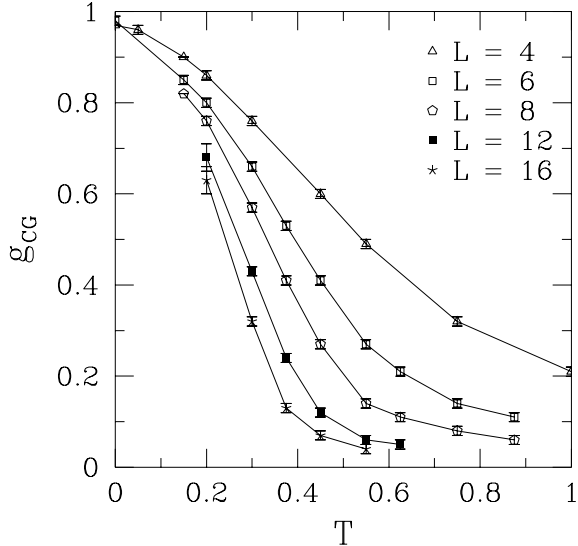}
\caption{Results for the Binder moment ratio $g_{\scriptscriptstyle CG}$,
defined in Eq.~(\protect\ref{g-def}), for different
sizes and temperatures. The lines are guides to the eye.}
\label{fig1}
\end{figure}

\begin{figure}
\epsfxsize=\columnwidth\epsfbox{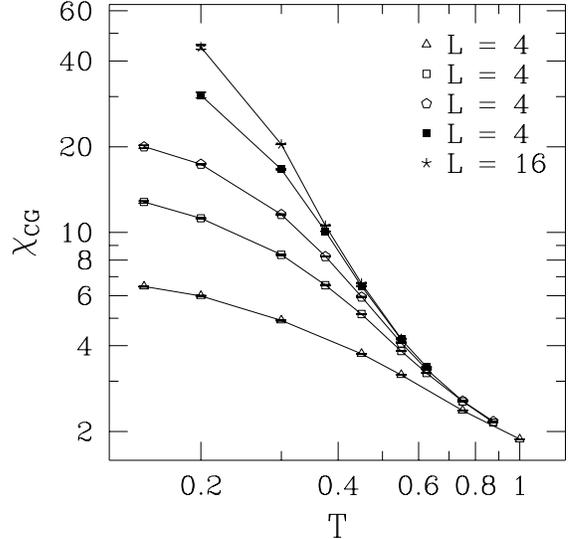}
\caption{A log-log plot of results for the chiral glass susceptibility
$\chi_{\scriptscriptstyle CG}$, defined in Eq.~(\protect\ref{chi-def}),
for different sizes and temperatures.
The lines are guides to the eye.}
\label{fig2}
\end{figure}

The results for $g_{\scriptscriptstyle CG}$ as a function of temperature
for different sizes are shown in Fig.~\ref{fig1}
and the corresponding results for $\chi_{\scriptscriptstyle CG}$
are in Fig.~\ref{fig2}. 
The points at $T = 0$ are obtained by exact enumeration of all the
states.

We also calculated the ground state degeneracy for L = 4 and L = 6 
and found that there is a small non-zero degeneracy, which leads to 
$g_{\scriptscriptstyle CG}$ being slightly below 1 at $T=0$. However,
$g_{\scriptscriptstyle CG}$ increases
slightly with increasing size, so we expect
that $g_{\scriptscriptstyle CG}$ will be unity at $T=0$ in the
thermodynamic limit. 
One should note here that Ray and Moore\cite{rm}, who worked in the phase
representation, also found that $g_{\scriptscriptstyle CG}$
is not very sensitive to the ground state degeneracy, and seems to
extrapolate to unity.


In order to do a finite size scaling analysis for $g_{\scriptscriptstyle
CG}$ it is convenient to incorporate one trivial correction to scaling.
As defined in Eq.~(\ref{g-def}), $g_{\scriptscriptstyle CG}$ tends to
$1/N$ (rather than 0) as $T \to \infty$,
which is not completely negligible for the
sizes studied here. We therefore consider the following quantity
\begin{equation}
\label{g_prime}
g^\prime_{\scriptscriptstyle CG} \equiv
{{N g_{\scriptscriptstyle CG} - 1} \over {N - 1}}
= \tilde{g}_{\scriptscriptstyle CG}(L^{1/\nu_{_{\!CG}}} T)\ ,
\end{equation}
which does vanish at high temperatures for finite $N$ (and tends to
unity as $T \to 0$ if $g_{\scriptscriptstyle CG}$ does).

\begin{figure}
\epsfxsize=\columnwidth\epsfbox{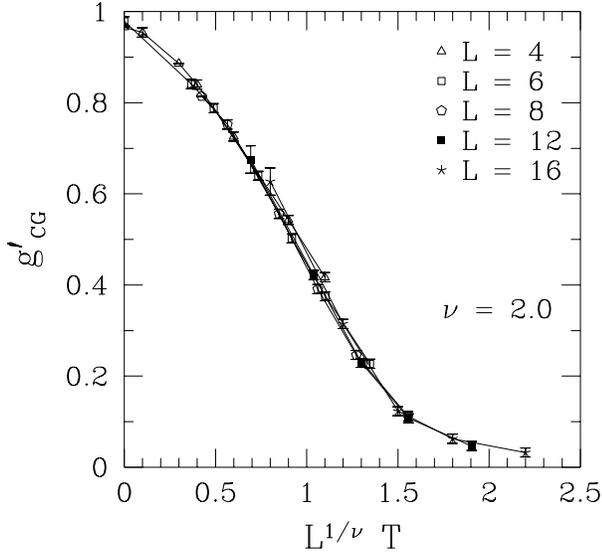}
\caption{A scaling plot of the data in Fig.~\protect\ref{fig1} according
to Eq.~(\protect\ref{g_prime}).
Data for $T \le 0.55$ has been included.
}
\label{fig3}
\end{figure}

Fig.~\ref{fig3} shows our finite size scaling 
plot for $g^\prime_{\scriptscriptstyle CG}$, including data for 
$T \le 0.55$. Data at higher temperatures did
not scale well and is presumably not in the scaling region. 
It is clear that good data
collapse obtains with $\nu_{\scriptscriptstyle CG} = 2.0$. 
We tried scaling with other values of
$\nu_{\scriptscriptstyle CG}$ and found that data collapse gets 
worse both for $\nu_{\scriptscriptstyle CG} > 2.0$
and for $\nu_{\scriptscriptstyle CG} < 2.0$. Trying various values
for $\nu_{\scriptscriptstyle CG}$
in this way we estimate $\nu_{\scriptscriptstyle CG} = 2.0 \pm 0.2$
from the data for $g_{\scriptscriptstyle CG}$ in Fig.~\ref{fig3}.

Just at it is convenient to put in the correction to
$g_{\scriptscriptstyle CG}$ in Eq.~(\ref{g_prime})
in order that it varies between 1 and
0 as the temperature changed, it is also useful to perform a
similar transformation for $\chi_{\scriptscriptstyle CG}$. Since
$\chi_{\scriptscriptstyle CG}$ tends to unity as $T \to 0$, we
subtract unity. Furthermore, our zero temperature data is consistent
with $\eta_{\scriptscriptstyle CG}$ very close to zero so, from now on,
we will assume that $\eta_{\scriptscriptstyle CG}=0$
and that $\chi_{\scriptscriptstyle CG} = N$ at $T=0$ in
the thermodynamic limit.
Hence we will analyze
\begin{equation}
\chi^\prime_{\scriptscriptstyle CG} \equiv {\chi_{\scriptscriptstyle CG}
- 1 \over N - 1}
= \tilde{\chi}_{\scriptscriptstyle CG}(L^{1/\nu_{_{\!CG}}} T) \ ,
\label{chi-prime}
\end{equation}
which varies between 1 and 0 as $T$ increases.

\begin{figure}
\epsfxsize=\columnwidth\epsfbox{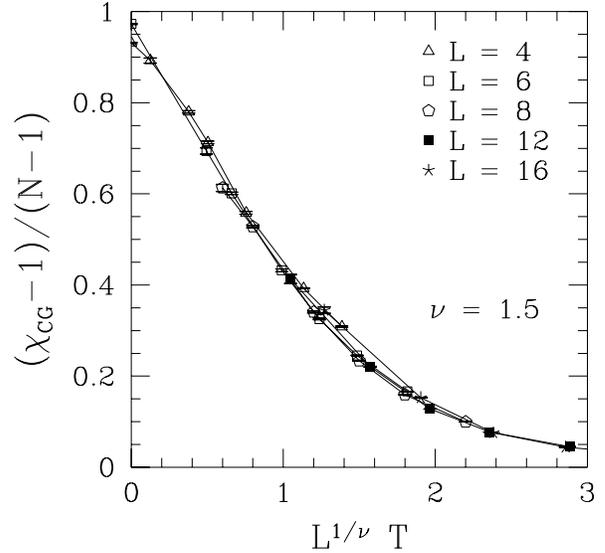}
\caption{A scaling plot of the data in Fig.~\protect\ref{fig2} according
to Eq.~(\protect\ref{chi-prime}) with $\nu_{\scriptscriptstyle
CG}= 1.5$.
The data used in the plot is for
$T \le 0.55$.
This plot {\em assumes} $\eta_{\scriptscriptstyle CG} = 0$, which
is reasonable
since the $T=0$ results indicate that
$\eta_{\scriptscriptstyle CG}$ is close to, and probably
exactly equal to, zero.
}
\label{fig4}
\end{figure}

In Fig.~\ref{fig4} we show a finite size scaling plot for
$\chi^\prime_{\scriptscriptstyle CG}$, including data for
$T \le 0.55$.
The best fit is for $\nu_{\scriptscriptstyle CG} = 1.5$,
somewhat lower than that
obtained from $g_{\scriptscriptstyle CG}$. A similar difference
is also found in the three-dimensional Ising spin glass\cite{ky}.

Combining our exponent estimates from $g_{\scriptscriptstyle CG}$
and $\chi_{\scriptscriptstyle CG}$ we obtain
\begin{eqnarray}
\nu_{\scriptscriptstyle CG} & = & 1.8 \pm 0.3 \nonumber \\
\eta_{\scriptscriptstyle CG} & = & 0.0 \pm 0.2  \ .
\end{eqnarray}
These results indicate that the chiralities in the
two-dimensional XY spin glass order with a correlation length
exponent $\nu_{\scriptscriptstyle CG}$ which is different from
the spin glass correlation length exponent, for which earlier
work\cite{rm,jy} found
$\nu_{\scriptscriptstyle SG} \simeq 1$. This
conclusion is in
agreement with earlier results\cite{kt,rm}.
For the future it
would be useful to study the {\em three\/}-dimensional
XY spin glass in the
vortex representation to understand whether there is indeed a finite
temperature chiral-glass phase as found in the work of Kawamura\cite{k}.

One of us (H.B)
would like to thank Tanya Kurosky for useful discussions. This
work is supported by the NSF DMR--9411964.

\end{document}